\documentclass[12pt]{article}
\usepackage{newinutile}
\usepackage{latexsym,amsmath,amsfonts,amsthm,amssymb,bbm,fdsymbol}
\usepackage{epsfig,graphics,color,calc,graphicx}
\usepackage{subfigure}
\usepackage{mathrsfs}
\newcommand{\newatop}[2]{\genfrac{}{}{0pt}{}{#1}{#2}}
\newcommand{\rr}[1]{{\normalfont\textrm{#1}}}
\newcommand{\bb}[1]{{\mathbb{#1}}}

\newlength{\pecettawidth}
\begin{document}
%%% Font: commenta la riga che segue se vuoi i font standard
%\fontfamily{ppl}\selectfont
%\logo
\title{Trapping in bottlenecks: interplay between 
microscopic dynamics and large scale effects}

\author{Emilio N.M.\ Cirillo}
\email{emilio.cirillo@uniroma1.it}
\affiliation{Dipartimento di Scienze di Base e Applicate per l'Ingegneria, 
             Sapienza Universit\`a di Roma, 
             via A.\ Scarpa 16, I--00161, Roma, Italy.}
%\thanks{ENMC acknowledges Eurandom for the kind hospitality.}

\author{Matteo Colangeli}
\affiliation{Dipartimento di Ingegneria e Scienze dell'Informazione e 
Matematica, Universit\`a degli Studi dell'Aquila, via Vetoio, 
67100 L'Aquila, Italy.}
\email{matteo.colangeli@univaq.it}

\author{Adrian Muntean}
\affiliation{Department of Mathematics and Computer Science,
             Karlstad University, Sweden.}
\email{adrian.muntean@kau.se}

%\thanks{The authors thanks....}

\begin{abstract}
We investigate the appearance of trapping states in pedestrian flows
through bottlenecks as a result of the interplay between the geometry
of the system and the microscopic stochastic dynamics.
We model the flow trough a bottleneck via a Zero Range Process on a
one dimensional periodic lattice.
Particle are removed from the lattice sites with
rates proportional to the local occupation numbers.
The bottleneck is modelled by a particular site of the lattice where
the updating rate saturates to a constant value as soon as the local occupation number exceeds a fixed threshold.
We show that, for any finite value of such threshold, the stationary particle current saturates to the limiting bottleneck rate when the total
particle density in the system exceeds the bottleneck rate itself.
\end{abstract}

%\pacs{}

\keywords{Pedestrian flows through bottlenecks; trapping; condensation;
stochastic modeling; interacting particle systems.}

\ams{90B06, 60K30, 82C22}

%\preprint{Appunti: \today}  

%\vfill
%\noindent
%\textbf{MSC2000:} 82B28; 82B44; 60K35.

\maketitle

\section{Introduction}
\label{s:introduzione} 
%\par\noindent
The effect of bottlenecks on a flow in a lane 
is relevant in many applied contexts 
such as traffic\cite{KR1996,YHT2006} and 
pedestrian\cite{KGS2006,LSZBZZ2014,HINT2003,DH2010,BBK2013} flows, 
motion in biological systems\cite{MC2017,BS2012},
but 
also in more abstract problems such as the study of the 
effect of blockage in stationary states 
\cite{JanLeb94,CCMpre2016,SLM15} or the effect of obstacles 
on two--dimensional particle flows\cite{CMKS2016,CKMSS2016,FPS2016}.

Bottlenecks are usually the effect of 
a local \emph{capacity reduction}
of the lane which can be due to different reasons
such as speed reduction, shrinkage of the lane, or decrease of the 
field governing the motion. In the case of pedestrian flows, which will 
be the application we shall focus on in the sequel, bottlenecks are 
typically due to the presence of a door or a corridor which 
shrinks the lane width inducing a direct capacity reduction. 
A huge amount of engineering research is ongoing on questions like
crowd evacuation, route choices, doors design and results are mostly
experimental, often not conclusive\cite{RKNPR2016,Metal2016}. 
A general agreement is lacking. As regards the
interplay between flow dynamics and door geometry, we discover that general
laws govern the structure of fundamental diagrams, and point out the presence of trapping (condensation) regimes. Interestingly, we shed also light on parameter regimes that prevent the onset of those trapping
states. Our findings, see Section~\ref{s:fondamentali},
are thus expected to have a relevant impact on
crowd management and building design, especially when big perturbations, 
e.g.\ due to fire, accidents, terrorist attacks, potentially occur in the 
flow of
pedestrians. In particular, we know exactly how the door size affects the
structure of the fundamental diagrams,
see Fig.~\ref{fig:fd000}.

Usually
the motion is affected by the reduced 
capacity of the lane only if the local density is 
sufficiently large, that is, namely, if the number of 
pedestrians moving through the reduced capacity region is high.
Think, for instance, to pedestrians walking through a door: 
if the number of people approaching the door per unit of time is 
low, such a capacity reduction will have no effect on the 
flow. On the other hand, if the approaching rate 
is high enough, pedestrians will be not able to pass through the door 
efficiently and the total flux will hence decrease.  
In other words, in these situations the capacity reduction will 
affect the flow only if the local density overcomes a certain
\emph{saturation threshold}. 

Thus, this blockage phenomenon depends essentially on two parameters: the 
saturation threshold and the reduced capacity. We shall 
develop a basic model to study this problem and, in this 
framework, we shall explain which of the two 
parameters actually controls the onset of the bottleneck.

We shall consider a one--dimensional asymmetric Zero Range Process (ZRP)
on the periodic lattice with updating rates proportional to the 
number of particles at each site, excepting on one
\emph{defect site}, modelling the reduced capacity portion 
of the lane, 
where the updating rate saturates to a value, called 
the \emph{saturated rate}, when the number of particles
exceeds a value, called the \emph{saturation threshold}.
When a site is updated a particle is moved forward or backward with 
a prescribed probability.
For the Zero Range models the idea of activation and saturation thresholds
was introduced in\cite{CCM-MMS16,CCM-CRM16}, where 
different interpretations, ranging from pedestrian dynamics to the 
thermodynamic theory of phase transitions, have been considered. 
We also mention that, in the recent literature, 
ZRP with modified blockage rules have also been studied in 
different frameworks, e.g.\ non--Markovian processes and traffic 
models\cite{HMS2009,HMS2012,CMH,KMH}.

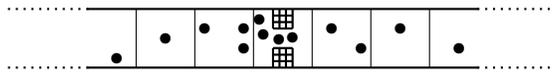
\begin{figure}
\begin{picture}(400,60)(-180,0)
\setlength{\unitlength}{.026cm}
%Left bottom
\thicklines
\put(-30,20){\line(1,0){200}}
\put(-30,50){\line(1,0){200}}
\thinlines
\put(55,20){\line(0,1){30}}
\put(85,20){\line(0,1){30}}
\put(25,20){\line(0,1){30}}
\put(115,20){\line(0,1){30}}
\put(-5,20){\line(0,1){30}}
\put(145,20){\line(0,1){30}}
%periodic conditions
\thicklines
\qbezier[10](170,20)(190,20)(210,20)
\qbezier[10](170,50)(190,50)(210,50)
\qbezier[10](-70,20)(-50,20)(-30,20)
\qbezier[10](-70,50)(-50,50)(-30,50)
%particelle
\put(10,35){\circle*{5}}
\put(-15,25){\circle*{5}}
\put(30,40){\circle*{5}}
\put(50,30){\circle*{5}}
\put(50,40){\circle*{5}}
\put(95,40){\circle*{5}}
\put(110,30){\circle*{5}}
\put(130,40){\circle*{5}}
\put(160,30){\circle*{5}}
\put(60,37){\circle*{5}}
\put(58,44.5){\circle*{5}}
\put(68,34.5){\circle*{5}}
\put(75,35.5){\circle*{5}}
%blockage
\thicklines
\put(65,20){\line(0,1){10}}
\put(65,30){\line(1,0){10}}
\put(75,30){\line(0,-1){10}}
\put(65,23.3){\line(1,0){10}}
\put(65,26.6){\line(1,0){10}}
\put(68.3,20){\line(0,1){10}}
\put(71.6,20){\line(0,1){10}}

\put(65,50){\line(0,-1){10}}
\put(65,40){\line(1,0){10}}
\put(75,40){\line(0,1){10}}
\put(65,43.3){\line(1,0){10}}
\put(65,46.6){\line(1,0){10}}
\put(68.3,40){\line(0,1){10}}
\put(71.6,40){\line(0,1){10}}
\end{picture}
\caption{Sketch of pedestrians moving on the lane.
The cell at the center is the one in which walkers experience the blockage 
and corresponds to defect site of the lattice model.
In the lattice model each cell is lumped to a single site.
}
\label{f:pede}
\end{figure}

This model can be thought of as a basic model for pedestrians on a lane
with a width shrinkage due to a corridor or a door.
The lane is partitioned in squared cells, each cell is a site of the ZRP
model and the defect site models the cell where 
the blockage takes place, see Fig.~\ref{f:pede}. 
The number of particles at a site is the number of 
pedestrians in the corresponding cell. 
Pedestrians moving in a given direction can be modelled by assuming 
that the particles can only jump forward (or, equivalently, backward). 
Pedestrians that can move back and forth on the lane 
can be modelled by assuming 
that the probabilities for a particle to move backward or forward on the 
lattice are both different from zero. 
Using a stochastic model, rather than a deterministic one,
allows to take into account the effects of density fluctuations 
in the pedestrian flow (real walkers do not move all at the same 
instant). As it will be explained in the following 
section, these fluctuations seem to induce a reduced flow
even in a setup in which the total pedestrian density on the lane would not 
justify it. 

Exploiting the theory of condensation for 
ZRP\cite{Evans00,EvansHanney05,GL2012,CG2015,GS2008}
we shall 
prove that, in such a setup, the parameter controlling the 
bottleneck onset is the saturated rate. More precisely, we shall 
see that, provided the total density of the system is large with respect 
to the saturated rate, the system exhibits a condensed state 
with a reduced flow proportional to the saturated rate. 
It is worth noting that the model studied in this paper
is an example of ZRP with updating rates not decreasing 
with the number of particles and exhibiting a condensed state induced by 
a local inhomogeneity in the updating rates. 
Our model should be compared with the one studied 
in\cite[Section~5.2]{EvansHanney05} where all the site updating 
rates are set equal to one except for one site which is updated at a  
lower constant rate. This peculiar behavior at one site is sufficient 
for the condensed state to appear, provided that the total density is 
large enough with respect to the lowered updating rate. 
In our model, on the contrary, the updating rate is proportional to the 
local occupation number for any site except the defect 
site, where it saturates to a constant value (possibly even much larger than one!). In the sequel we shall prove that also this updating mechanism may lead to 
condensation, provided the total density is large enough with respect to the 
saturated constant rate. 

This result, in terms of the pedestrian interpretation, can be 
rephrased as follows. The pedestrian current is decreased by the 
capacity reduction on the lane if pedestrian total 
density is large compared to the saturated rate which pedestrians 
experience in the region of the lane with capacity reduction. 
Moreover, in this regime the pedestrian flux is proportional to the 
saturated rate, which, in the case of a capacity reduction caused by the presence of a door,
it is reasonable to assume to be proportional to the door width. Note that this result is indeed 
found in different experimental setups\cite{YHT2006,KGS2006}.
We shall also unravel the main features of the fundamental diagram associated 
to our model, in which we plot the local speed as a function of the 
local particle concentration. In particular, we find results showing striking agreement
with a set of experimental data measured in a situation similar to the one considered in our abstract setup\cite{DH2005}.

The paper is organized as follows. In Section~\ref{s:modello}
we define the model and discuss the results. 
Section~\ref{s:analitico} is devoted to the analytical study of the model.
In Section~\ref{s:fondamentali} we discuss our results in view of the 
pedestrian flow interpretation of the model. 
%Finally, 
%in Section~\ref{s:conclusioni} we briefly summarizes our conclusions. 

\section{Model and results}
\label{s:modello} 
%\par\noindent
To define the ZRP to be studied in this paper we borrow the notation 
from\cite{EvansHanney05}.
We consider the positive integers $L,N$, 
the finite torus $\Lambda=\{1,\dots,L\}$, and 
the finite \emph{state} or \emph{configuration space}
$\Omega_{L,N}$
made of the states 
$n=(n_1,\dots,n_L)\in\{0,\dots,N\}^\Lambda$
such that 
$\sum_{x=1}^Ln_x=N$.
Given
$n\in\Omega_{L,N}$
the integer $n_x$ is called \emph{number of particle}
at site $x\in\Lambda$ in the \emph{state} or \emph{configuration}
$n$.
The integer $1\le T\le N$ and the real $c>0$ 
are respectively called \emph{saturation threshold}
and \emph{saturated rate}.
For any site $x\in\Lambda$,  
the hopping rate $u_x:\bb{N}\to\bb{R}_+$ is defined as follows:
$u_x(0)=0$ for $x=1,\dots,L$, 
$u_1(k)=k$ for $1\le k\le T$ and 
$u_1(k)=c$ for $T+1\le k\le N$,
and 
$u_x(k)=k$ for $x=2,\dots,L$ and $1\le k\le N$.
The ZRP considered in this context 
is the continuous time Markov process $n(t)\in\Omega_{L,N}$, $t\ge0$,
such that each site 
$x$ is updated with rate $u_x(n_x(t))$ and, 
once a site $x$ is chosen, a particle is moved 
to the neighboring site $x+1$ with \emph{forward hopping probability} 
$1/2<p\le1$ and 
to the neighboring site $x-1$ with probability $0\le1-p<1/2$
(recall that periodic boundary conditions are imposed).
Note also that when $T=N$ the model is
equivalent to independent particles random walk. 

It can be proven, see e.g.\cite[equations~(2) and (15)]{EvansHanney05},
that the \emph{invariant} or \emph{stationary measure}
of the ZRP process is 
\begin{equation}
\label{mod020}
\mu_{L,N}(n)
=
\frac{1}{Z_{L,N}}
\prod_{\newatop{x=1,\dots,L:}{n_x\neq0}}
\frac{1}{u_x(1)\cdots u_x(n_x)}
\end{equation}
for any $n\in\Omega_{L,N}$, where the
\emph{partition function}
$Z_{L,N}$ is the normalization constant
\begin{equation}
\label{Zthr}
Z_{L,N}
=\sum_{n\in\Omega_{L,N}}
\prod_{\newatop{x=1,\dots,L:}{n_x\neq0}}
\frac{1}{u_x(1)\cdots u_x(n_x)}
\;\;.
\end{equation}

The \emph{stationary current} $J_{L,N}$ represents the 
average number of particles crossing a bond between two given
sites in unit time and is defined by
$\mu_{L,N}[pu_x-(1-p)u_{x+1}]$. 
Since periodic boundary conditions are imposed,
the current does not depend on the chosen bond and is 
given by 
\begin{equation}
\label{mod050}
J_{L,N}
=(2p-1)\mu_{L,N}[u_x]
%=
%\frac{Z_{L,N-1}}{Z_{L,N}}
%Z_{L,N-1}/Z_{L,N}
\;\;.
\end{equation}
%The first equality defines the current, whereas the second one is proven 
%in\cite[equation~(11)]{EvansHanney05}. 
Other relevant quantities are the stationary 
mean occupation numbers
given by 
$m_{x,L,N}=\sum_{n\in\Omega_{L,N}}n_x\mu_{L,N}(n_x)$, for any $x=1,\dots,L$, 
and 
the stationary 
\emph{particle fraction} at the defect site 
$\nu_{L,N}=m_{1,L,N}/N$.

The main results discussed in the sequel 
will be deduced in the thermodynamic limit 
$N,L\to\infty$, with $N/L=\rho$ being the 
\emph{total constant density}.
When discussing the thermodynamic limit,  we shall drop the 
subscripts $L$ and $N$ from the notation 
and write $J$, $m_x$, and $\nu$ for the stationary current,
the mean occupation number, and the particle 
fraction at site $1$, respectively.

In the next section we show that 
both the particle fraction (at the defect site) and 
the stationary current suddenly change when the total density 
crosses the line $\rho=c$. 
In particular, the particle fraction at the defect site is zero for 
$\rho<c$ and positive for $\rho>c$; thus, 
the stationary states for $\rho<c$ and $\rho>c$ are
respectively called \emph{fluid} and \emph{condensed}.
Detailed results are listed in 
Table~\ref{t:mod000}.

\begin{table}
\begin{center}
\begin{tabular}{c|c|c|c|c}
\hline\hline
 & $J$ & $m_1$ & $\nu$ & $m_x,\,x\neq1$\\ 
\hline
$\rho<c$ & $(2p-1)\rho$ & $\rho$ & $0$ & $\rho$\\
\hline
$\rho>c$ & $(2p-1)c$ 
& $\stackrel{L\to\infty}{\sim}(\rho-c)L+c$ & $(\rho-c)/\rho$ & $c$\\
%$\rho>c$ & $(2p-1)c$ & $(\rho-c)L$ & $(\rho-c)/\rho$ & $c$\\
\hline\hline
\end{tabular}
\end{center}
\caption{Stationary current, mean occupation number at the defect site, 
particle fraction at the defect site, and mean occupation number at 
the regular sites in the fluid ($\rho<c$) and 
condensed ($\rho>c$) state in the thermodynamic limit. 
For $m_1$ in the condensed state the large $L$ behavior is reported.}
\label{t:mod000}
\end{table}

In the fluid state particles are distributed uniformly throughout the 
system with mean occupation number $\rho$. Since $\rho<c$,
sites are typically updated with rate $\rho$ and, consequently,
the current is equal to $(2p-1)\rho$. 
In the condensed state, instead, the occupation number at the 
defect site is proportional to $N$, hence, for $L$ sufficiently large, it will
exceed the saturation threshold $T$. We then have that the rate 
at which particles depart from such a site is $c$: this explain the 
value $(2p-1)c$ for the current. Moreover, since the current 
must be the same throughout the system, the rates at which the regular sites 
are updated attain the same value $c$. This explains why the mean occupation 
number at the regular sites is equal to $c$.

\begin{figure}
\begin{picture}(80,150)(0,0)
\includegraphics[width=0.5\textwidth]{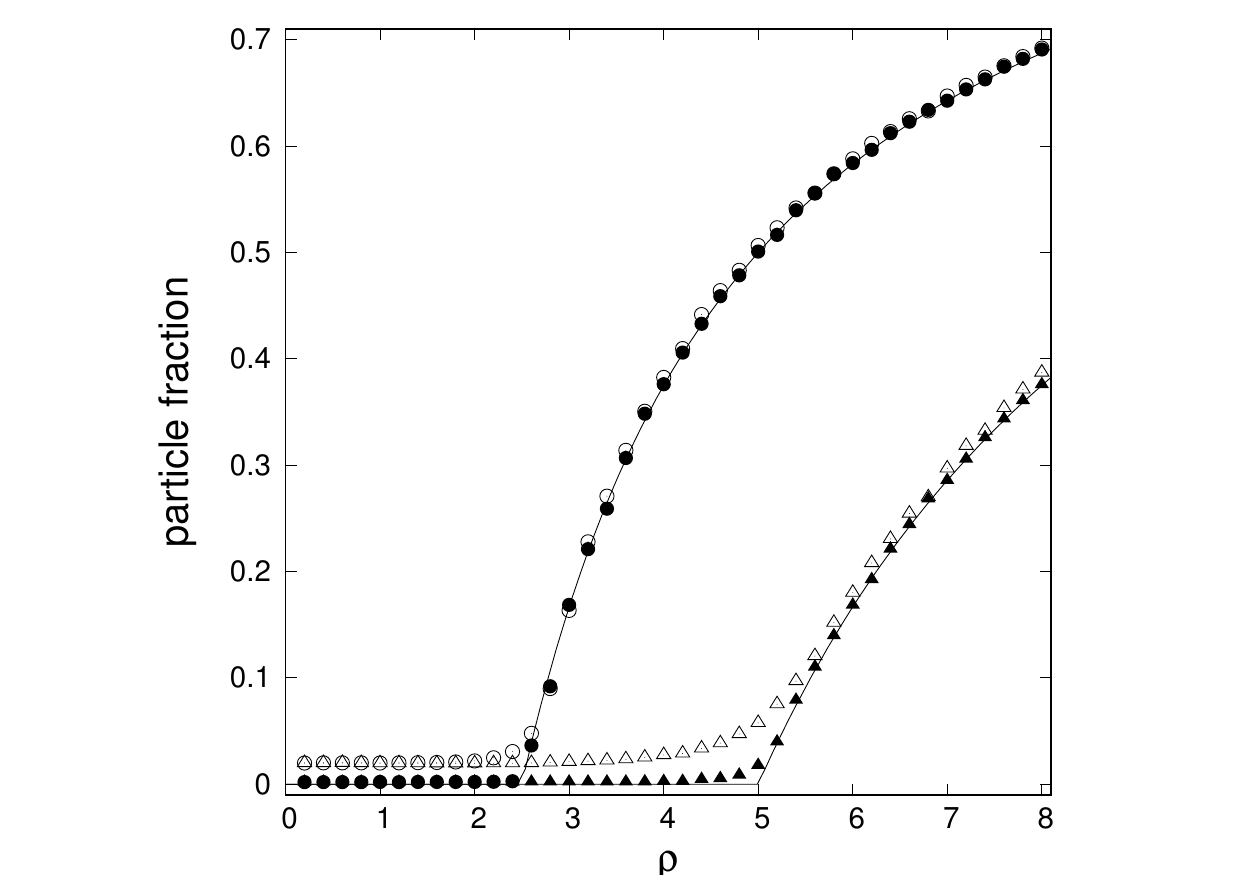}
\hskip 0.1 cm
\includegraphics[width=0.5\textwidth]{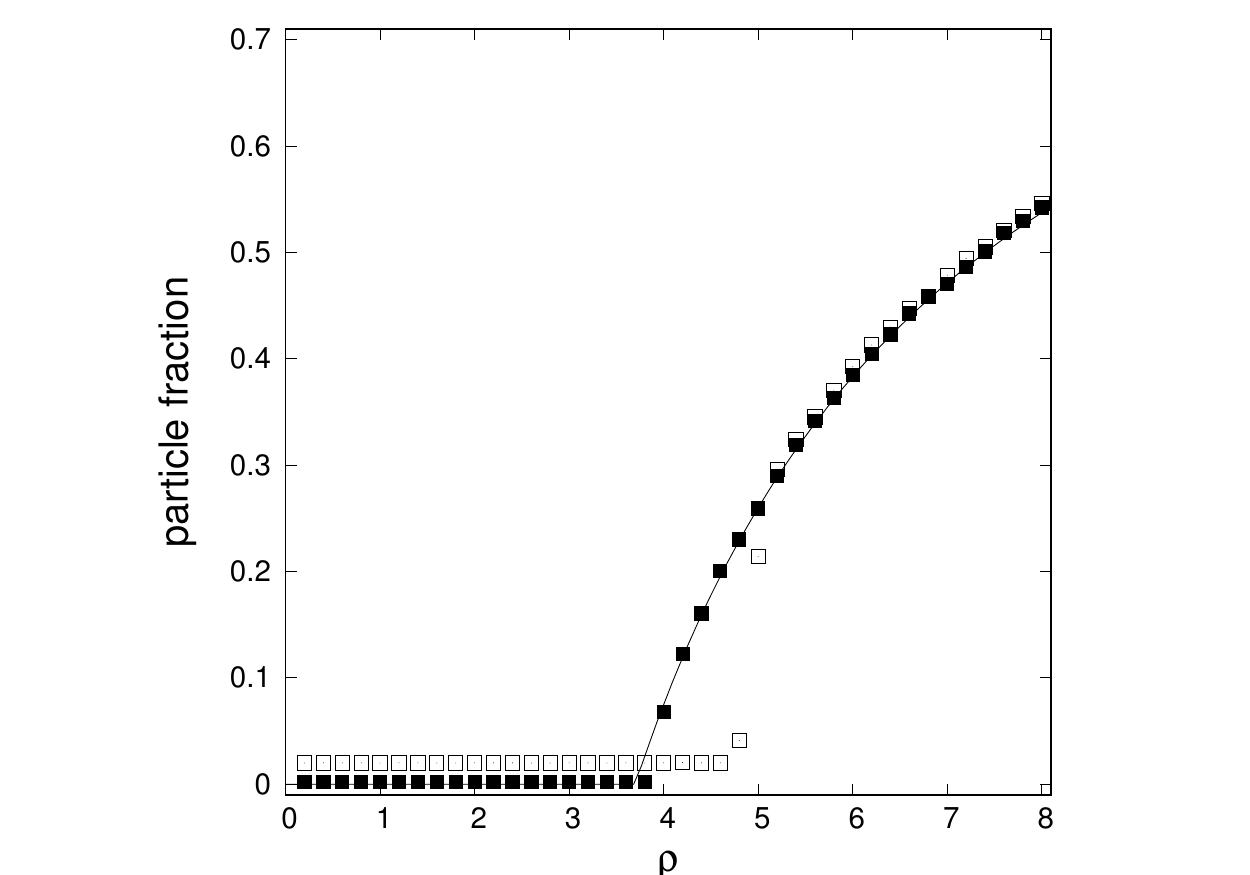}
\end{picture}
\caption{Stationary particle fraction $\nu$ (at the defect site) vs.\ $\rho$.
Open and solid symbols are the Monte Carlo prediction 
for $L=50$ and $L=500$, respectively. 
The forward hopping probability is $p=1$.
Circles and triangles refer, respectively, to
$T=6$ and $c=2.5$ 
($\smallcircle$ and $\smallblackcircle$)
and
$T=3$ and $c=5$ 
($\smalltriangleup$ and $\smallblacktriangleup$).
Squares refer to
$T=15$ and $c=3.7$ 
Solid lines are the theoretical predictions in Table~\ref{t:mod000}.}
\label{fig:thr1}
\end{figure}

\begin{figure}
\begin{picture}(80,150)(0,0)
\includegraphics[width=0.5\textwidth]{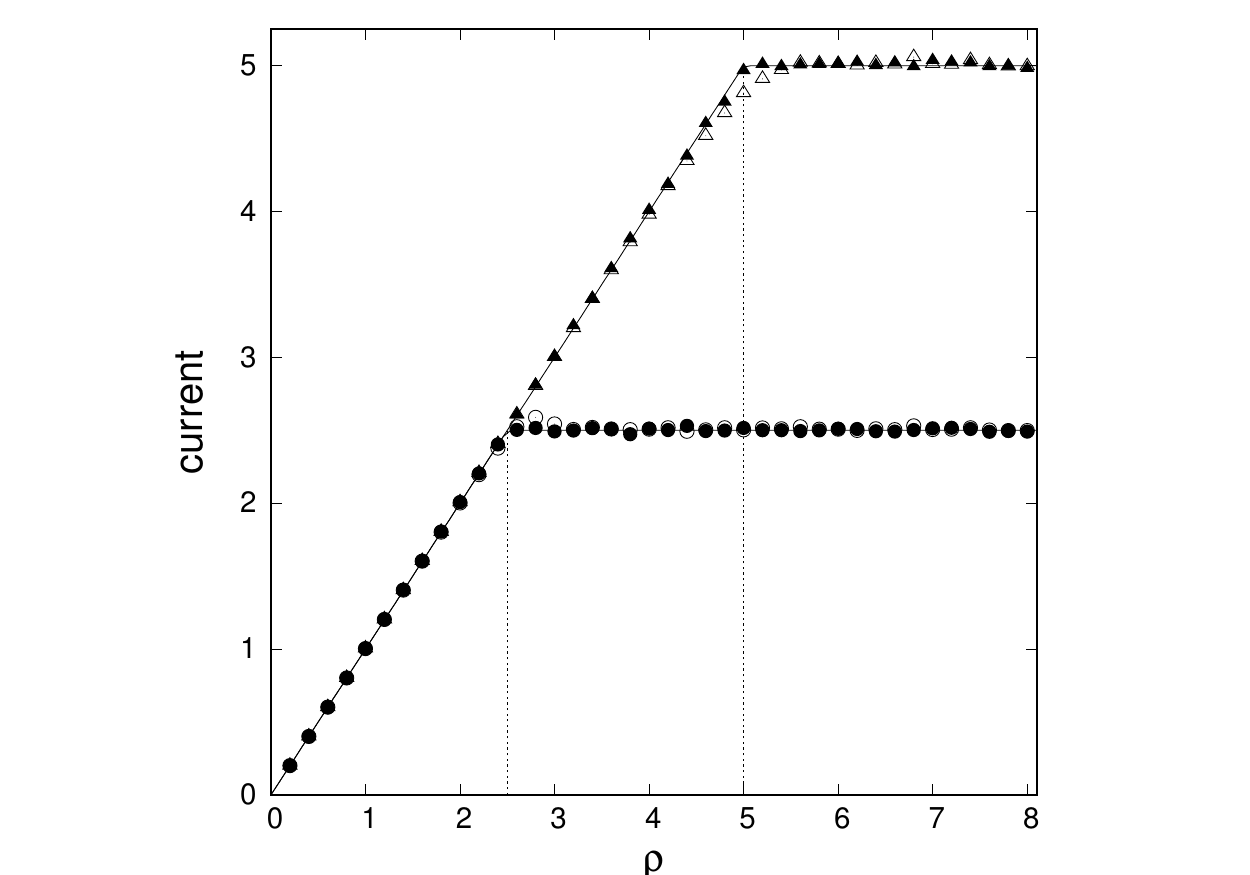}
\hskip 0.1 cm
\includegraphics[width=0.5\textwidth]{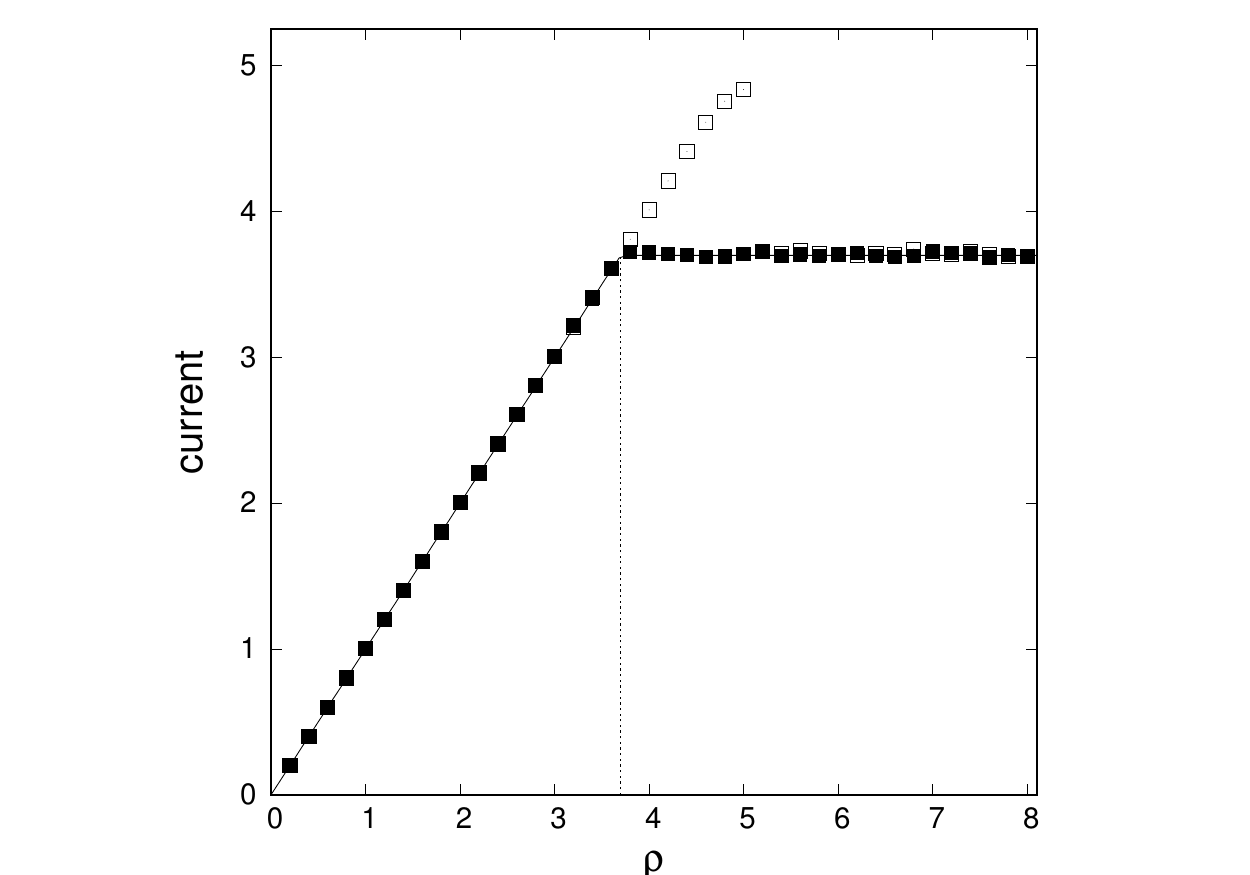}
\end{picture}
\caption{Stationary current vs.\ $\rho$.
Symbols are as in Fig.~\ref{fig:thr1}. 
Dotted lines indicate, respectively, the values of the total density corresponding to
the saturated rate $c=2.5$ and $c=5.0$ (left panel) and $c=3.7$ (right panel).  }
\label{fig:thr2}
\end{figure}

\begin{figure}
\begin{picture}(80,150)(0,0)
\includegraphics[width=0.5\textwidth]{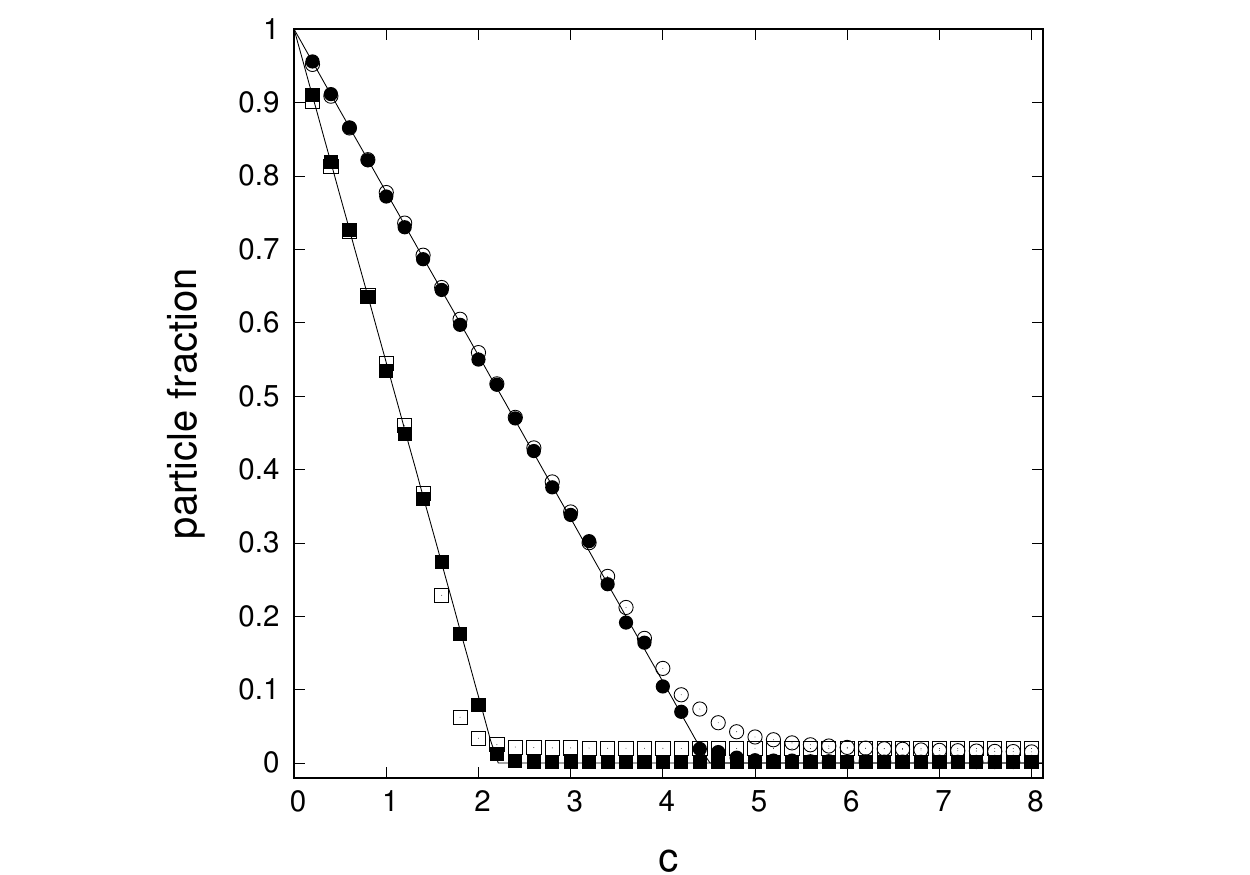}
\hskip 0.1 cm
\includegraphics[width=0.5\textwidth]{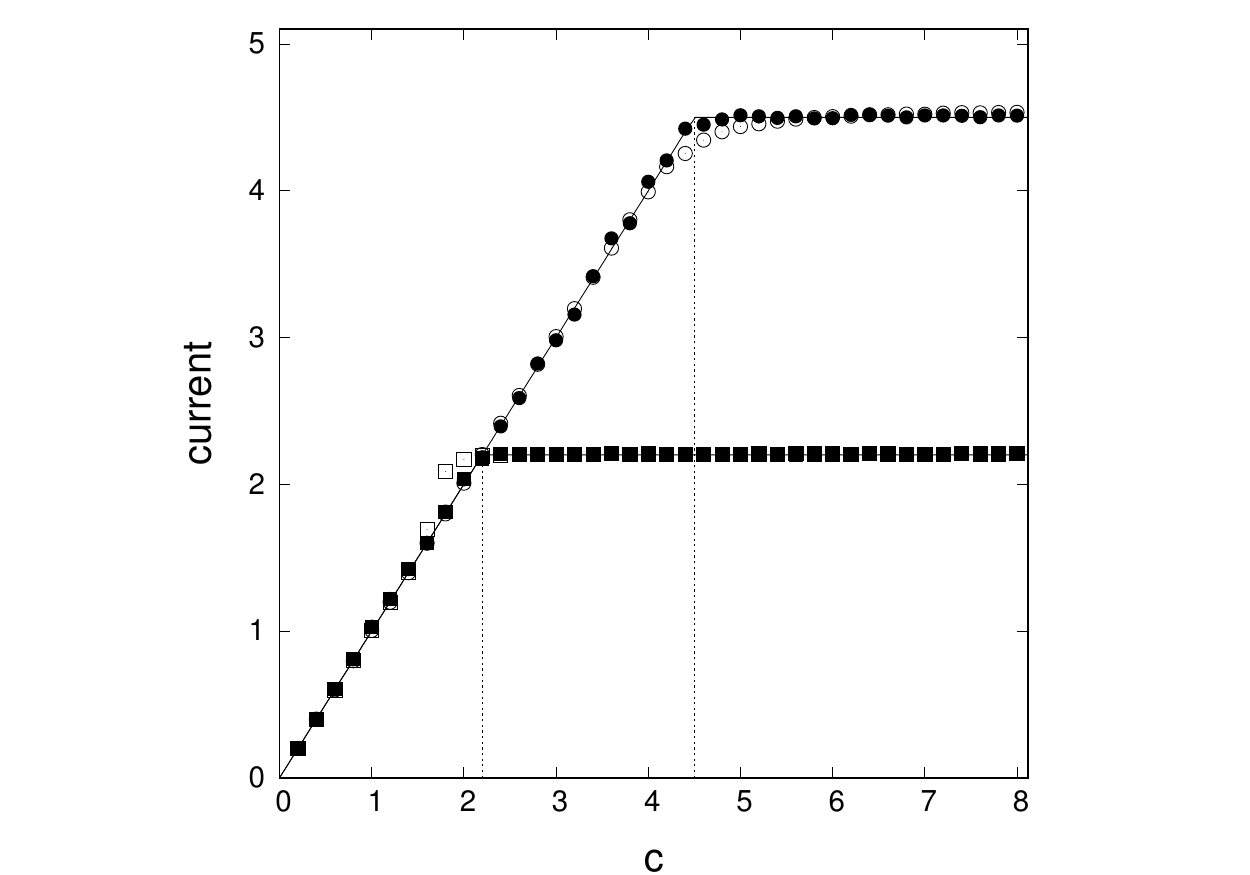}
\end{picture}
\caption{Particle fraction at the defect site (left panel) 
and stationary current (right panel) vs.\ the saturated rate $c$.
Open and solid symbols are the Monte Carlo prediction 
for $L=50$ and $L=500$, respectively. 
The forward hopping probability is $p=1$.
Circles and squares refer, respectively, to
$T=3$ and $\rho=4.5$ 
($\smallcircle$ and $\smallblackcircle$)
and
$T=7$ and $\rho=2.2$ 
($\smallsquare$ and $\smallblacksquare$).
Solid lines are the theoretical predictions in Table~\ref{t:mod000}, whereas the dotted lines indicate, respectively, the values of 
the saturated rate $c=2.2$ and $c=4.5$.}
\label{fig:thr3}
\end{figure}

Analytical results are plotted in 
figures~\ref{fig:thr1}--\ref{fig:thr3}
together with the results of Monte Carlo simulations performed as follows:
call $n(t)$ the configuration at time $t$, (i)
a number $\tau$ is picked up at random with
exponential distribution of
parameter $\sum_{x=1}^Lu_x(n_x(t))$ and time is update to
$t+\tau$, (ii)
a site is chosen at random on the lattice with probability
$u_x(n_x(t))/\sum_{x=1}^Lu_x(n_x(t))$, and (iii) a particle
is then moved from that site to the neighboring site on the right.
The results shown in the figures reveal a very good match between 
the analytical prediction and the numerical measures, moreover, 
the agreement improves when the lattice size $L$ increases. 

Figures~\ref{fig:thr1} and \ref{fig:thr2} show Monte Carlo and 
analytical results when the saturation threshold and the saturated 
rate are kept fixed whereas the total density is varied 
from $0.2$ to $8$. Three cases are considered, namely, 
$T=3$, $c=5$, and $p=1$ (triangles), 
$T=6$, $c=2.5$, and $p=1$ (circles),
and
$T=15$, $c=3.7$, and $p=1$ (squares).
Results show neatly that the system remains in the fluid 
state until the total density exceeds the saturated rate $c$.
Indeed, at $\rho=c$ the particle fraction, equal to zero in 
the fluid state, starts to increase. Note that simulations were 
stopped at $\rho=8$; a further increase in $\rho$ would correspond to a particle fraction at the defect site saturating to one.
The current in the fluid state increases linearly with $\rho$, as 
$\rho$ is the average updating rate throughout the system. 
On the other hand, it is constantly equal to the saturated rate $c$ in 
the condensed state.

For $L=500$ the match between the numerical measurement and the analytical 
computation, valid in the thermodynamic limit, is striking. 
At $L=50$ finite size effects are visible. 
In particular we note that in the case with large threshold, i.e., $T=15$, 
the stationary state switches from the fluid to the condensed one 
when the total density overcomes the value corresponding to $c=3.7$. 
On the other hand, in the case $L=50$ the stationary state persists in the 
wrong fluid phase until the total density reaches approximately the value $5$. 
This effect can be explained as follows: 
in Section~\ref{s:analitico} we shall see that the two 
stationary states are associated with two minima of the function 
$I(k)$ introduced below equation \eqref{ana050}. 
At finite volume the function $I$ exhibits two minima, whereas 
in the thermodynamic limit only one of the two minima survives.
Moreover, in the particular case shown in the picture, at $L=50$, 
the wrong fluid minimum is deeper than the one corresponding to the 
condensed one: for this reason, the systems appears to be trapped in the fluid 
state. This effect is no longer visible for small values of the threshold, cf. the data referring to the cases $T=3$ and $T=6$.  

Figure~\ref{fig:thr3} shows Monte Carlo and 
analytical results when the saturation threshold and the total density
are kept fixed whereas the saturated rated is varied 
from $0.2$ to $8$. Two cases are considered, namely, 
$T=7$, $\rho=2.2$, and $p=1$ (squares) and $T=3$, $\rho=4.5$, 
and $p=1$ (circles). 
Results show neatly that the system remains in the condensed
state until the saturated rate $c$ exceeds the total density $\rho$.
Indeed, at $c=\rho$ the particle fraction, linearly decreasing with $c$ in 
the condensed state, becomes constantly equal to zero. 
The current in the condensed state increases linearly with $c$ and
stays constantly equal to the total density $\rho$ in 
the fluid state.

Numerical results confirm the theoretical prediction 
in Table~\ref{t:mod000} on the phase diagram of the model: 
fluid state for $\rho<c$ and condensed state for $\rho>c$. 
The phase diagram can be justified intuitively imagining to prepare 
the system in the fluid state and trying to guess the consequent 
evolution.
Indeed, 
the behavior for $\rho>\max\{T,c\}$
is rather intuitive: initially the typical 
number of particles at each site would be above the threshold and
particles would leave the defect site at rate $c$ and the 
regular sites at rate $\rho>c$, so that eventually the system would reach 
the condensed state.
The behavior in the case $\rho<\min\{T,c\}$ is somehow opposite: in the fluid 
state the typical number $\rho$ of particles at each site would be smaller than
$T$ and $c$. Particles would leave each site of the lattice with rate $\rho$ 
and the system would remain in the fluid state. 
The case $T<\rho<c$ is more subtle: particles would leave the 
defect site at rate $c$ and the regular sites at rate $\rho<c$. 
The rate limitation at the defect site is not effective and the 
system remains in the fluid state. 
The case $c<\rho<T$ is the most interesting one: since $\rho<T$, in the fluid 
state each site is initially left by particles at the same rate $\rho$. 
But if a random fluctuation increased the number of particles 
at the defect site to a value larger than $T$, particles would then start 
to leave such a site at a rate smaller than $\rho$ (the rate at which
particles leave the regular sites in the fluid state), which would thus induce the 
transition to the condensed state. Being $T$ finite, 
we expect that the probability for such a fluctuation is so large 
to justify that the stationary state is the condensed one. 

\section{Analytical solution}
\label{s:analitico} 
%\par\noindent
The model will be studied by using techniques
similar to those developed in\cite[Section~5.2]{EvansHanney05} 
and\cite{CCMpre2016}.
We first recall the expression 
\begin{equation}
\label{ana000}
J_{L,N}
=
(2p-1)\frac{Z_{L,N-1}}{Z_{L,N}}
\end{equation}
for the current
proven in\cite[equation~(11)]{EvansHanney05}. 
Thus, our strategy will be the following: we shall first 
compute the partition function $Z_{L,N}$ and use \eqref{ana000}
to deduce the current. 
Then, exploiting \eqref{mod050} and the definition of the updating 
rates at the regular sites, we shall compute
\begin{equation}
\label{ana010}
m_{x,L,N}
=\frac{1}{2p-1}\,J_{L,N}
\end{equation}
for $x=2,\dots,L$.
Finally, the trivial equality, 
\begin{equation}
\label{ana012}
m_{1,L,N}=N-(L-1)m_{x,L,N} 
\end{equation}
with $x$ any site different from the defect one, will provide
\begin{equation}
\label{ana015}
\nu
=1-\frac{1}{\rho}\,m_{x,L,N}
\;\;.
\end{equation}

Simple algebra and the use of the multinomial 
theorem, see, e.g.\cite[equation~(3.35)]{Nelson1995} allow to 
rewrite the partition function \eqref{Zthr} as
\begin{equation}
\label{ana020}
Z_{L,N}
=
Z^{(1)}_{L,N}
+
Z^{(2)}_{L,N}
\end{equation}
with
\begin{equation}
\label{ana030}
Z^{(1)}_{L,N}
=
\sum_{k=0}^T\frac{1}{k!(N-k)!}(L-1)^{N-k}
\end{equation}
and
\begin{equation}
\label{ana040}
Z^{(2)}_{L,N}
=
\frac{c^T}{T!}
\sum_{k=T+1}^N\frac{1}{c^k(N-k)!}(L-1)^{N-k}
\end{equation}

In $Z^{(1)}_{L,N}$ the sum extends to the finite value $T$, thus the 
factorial can be approximated as 
\begin{equation}
\label{ana045}
(N-k)!
=
\frac{N!}{N(N-1)\cdots(N-k+1)}\approx\frac{N!}{N^k}
\end{equation}
yielding
\begin{equation}
\label{ana050}
Z^{(1)}_{L,N}
\approx
\frac{(L-1)^N}{N!}
\sum_{k=0}^T\frac{1}{k!}\,\rho^k
\end{equation}

The estimate of the sum in \eqref{ana040} is more delicate since 
the index $k$ can be arbitrarily large when the thermodynamic limit
is considered.
To evaluate the behavior of the partition function in the above 
limit, it is useful to introduce the function 
$I(k)$  by rewriting \eqref{ana040} as 
$Z^{(2)}_{L,N}=\sum_{k=0}^N\exp\{L I(k)\}$.
To understand where the maxima of $I(k)$ are located, 
we express $I(k+1)-I(k)$ as 
\begin{displaymath}
I(k+1)-I(k)
=
\frac{1}{L}
     \log\frac{N-k}{c(L-1)}
\;\;,
\end{displaymath}
which implies that 
$I(k+1)-I(k)>0$ if and only if 
$k<N-c(L-1)$.
Hence, for $\rho<c$ the function $I(k)$ attains its maximum 
value at $k=T+1$, whereas for $\rho>c$ the maximum is 
at $k^*=\lfloor(\rho-c)L\rfloor$.

\textit{Case $\rho<c$.\/}
The sum \eqref{ana040} is dominated by the first terms, hence, 
recall $T$ is finite, the factorial can be treated as in \eqref{ana045}. Thus, 
\begin{displaymath}
Z^{(2)}_{L,N}
\approx
\frac{c^T}{T!}\,\frac{(L-1)^N}{N!}
\sum_{k=T+1}^N\Big(\frac{\rho}{c}\Big)^k
\end{displaymath}
Performing the change of variables $h=k-(T+1)$ and extending the sum 
up to infinity we find 
\begin{equation}
\label{ana070}
Z^{(2)}_{L,N}
\approx
\frac{\rho^T}{T!}
\,\frac{(L-1)^N}{N!}
\,\frac{\rho/c}{1-\rho/c}
\;\;.
\end{equation}

Finally, using \eqref{ana000}, \eqref{ana050}, and \eqref{ana070} 
we get $J=(2p-1)\rho$. Moreover, \eqref{ana010} and \eqref{ana012}
yield $m_x=\rho$ for any $x=1,\dots,L$.

\textit{Case $\rho>c$.\/}
The sum \eqref{ana040} is dominated by the terms in an interval 
centered at $k^*=\lfloor(\rho-c)L\rfloor$.
The factorial in \eqref{ana040} can be approximated using the Stirling formula;
setting $x=k/L$ we have 
\begin{displaymath}
\frac{1}{c^k(N-k)!}(L-1)^{N-k}
=
\frac{1}{\sqrt{2\pi L(\rho-c)}}
e^{LF(x)}
\end{displaymath}
with
\begin{displaymath}
F(x)=
(\rho-x)-x\log c-(\rho-x)\log(\rho-x)
\;\;.
\end{displaymath}
Hence, the sum in \eqref{ana040} can be approximated 
as the following integral 
\begin{displaymath}
Z^{(2)}_{L,N}
=
\frac{c^T}{T!}\sqrt{\frac{L}{2\pi}}
\int_{(T+1)/L}^\rho\frac{1}{\sqrt{\rho-x}}\,e^{LF(x)}\,\rr{d}x
\end{displaymath}

The function $F(x)$ has obviously a maximum at $x^*=\rho-c$.
We can expand the exponent in Taylor series up to the second order 
and compute the Gaussian integral to get 
\begin{equation}
\label{ana080}
Z^{(2)}_{L,N}
\approx
\frac{c^T}{T!}\frac{1}{\sqrt{c}}
e^{Lc-N\log c}
\;\;.
\end{equation}

To compare \eqref{ana080} and \eqref{ana050}, we use the Stirling 
approximation to write 
$(L-1)^N/N!\approx\exp\{N-N\log\rho\}/\sqrt{2\pi N}$.
Since,
$Lc-N\log c> N-N\log\rho$, we have that 
\begin{equation}
\label{ana090}
Z_{L,N}
\approx
\frac{c^T}{T!}\frac{1}{\sqrt{c}}
e^{Lc-L\log c}
\bigg(1+\frac{Z^{(1)}_{L,N}}{Z^{(2)}_{L,N}}\bigg)
\end{equation}
with 
$Z^{(1)}_{L,N}/Z^{(2)}_{L,N}\to0$ in the thermodynamic limit. 

Finally, using \eqref{ana000} and \eqref{ana090} 
we get $J=(2p-1)c$. Moreover, \eqref{ana010} and \eqref{ana015}
yield $m_x=c$ for any $x=2,\dots,L$
and $\nu=(\rho-c)/\rho$.

\section{Discussion}
\label{s:fondamentali} 
\par\noindent
We discuss our results 
in view of the pedestrian flow interpretation proposed in 
Section~\ref{s:introduzione}
(see also Fig.~\ref{f:pede}), sticking to the case $p=1$.

From the point of view of pedestrian flows, in 
the ZRP model each regular site is left by a 
walker at rate equal to the number of people which are 
at that time at the site. This means that in a time of order one 
all the people at the site will abandon it, hence, the local 
current is equal to the occupation number of the cell. 
For the same reason, if the defect site occupation number 
is small (smaller than the saturation threshold),
the outgoing current is equal to the defect cell 
occupation number as well.
On the other hand, if the defect site occupation number 
is large, the number of walkers leaving such a site in a unit of time 
is equal to the constant value $c$ (the saturated rate) and 
hence the local current is equal to $c$.

Thus, in case the total density $\rho$ is smaller than $c$, one can guess 
that the stationary state has the walkers distributed 
uniformly at the sites of the lattice. Indeed, at each site the 
local current would be equal to $\rho$. 
This is the \textit{fluid state}.
On the other hand, if $\rho>c$, it can happen that, 
at a certain time, a bunch of people larger than the 
saturation threshold reaches simultaneously 
the defect cell. This would reduce, on that site, the local current to $c$,
giving then rise to a stationary state characterized by an accumulation of walkers 
at the blocked site
and a stationary current equal to $c$.
This is what we call \textit{condensed} or \textit{trapping state}.
This is a remarkable phenomenon.
Indeed, even in the case in which 
the capacity reduction becomes effective at 
the defect site only above a certain saturation threshold larger than 
the saturation rate (note that we can compare these two quantities since 
the particle speed away from the bottleneck is one), 
provided $\rho>c$, 
the ``jammed" state can be induced by a local fluctuation larger than the 
saturation threshold,
even if the total density $\rho$ is smaller than the activation 
threshold $T$. 

Imagine $c$ to 
be proportional to the width of a door or corridor disturbing 
the pedestrian flow. 
Then the capacity reduction affects the pedestrian rate.
Essentially, 
the pedestrian flow appears to increase proportionally to the width 
of the door as observed in the experiments in\cite{YHT2006,KGS2006}.

A relevant question about pedestrians flow in presence of bottlenecks
is that of understanding the dependence of the pedestrian speed 
on the local density, namely, the so called \emph{fundamental diagram}.
We shall define the local speed of the particles 
as the ratio between the stationary current and the stationary 
occupation number. 
Since, away from the defect site, the current and the 
stationary occupation number are equal (see the results 
in Table~\ref{t:mod000} for $p=1$), the walkers
speed is one independently of the number 
of walkers at the site. This is quite obvious for our version of 
the ZRP model, since excepted for the defect site the site updating 
rate is proportional to the occupation number. 

\begin{figure}
\begin{picture}(80,150)(0,0)
\includegraphics[width=0.5\textwidth]{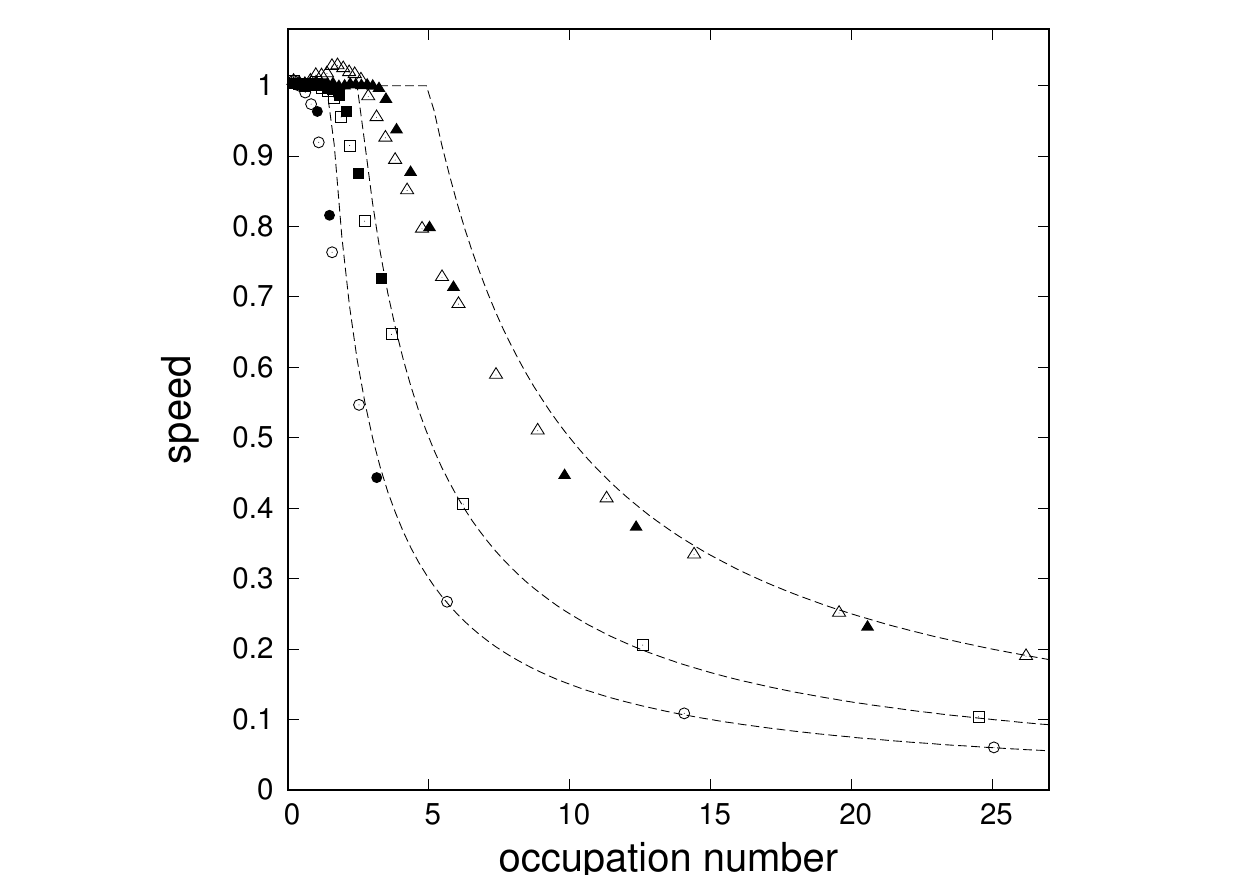}
\hskip 0.1 cm
\includegraphics[width=0.5\textwidth]{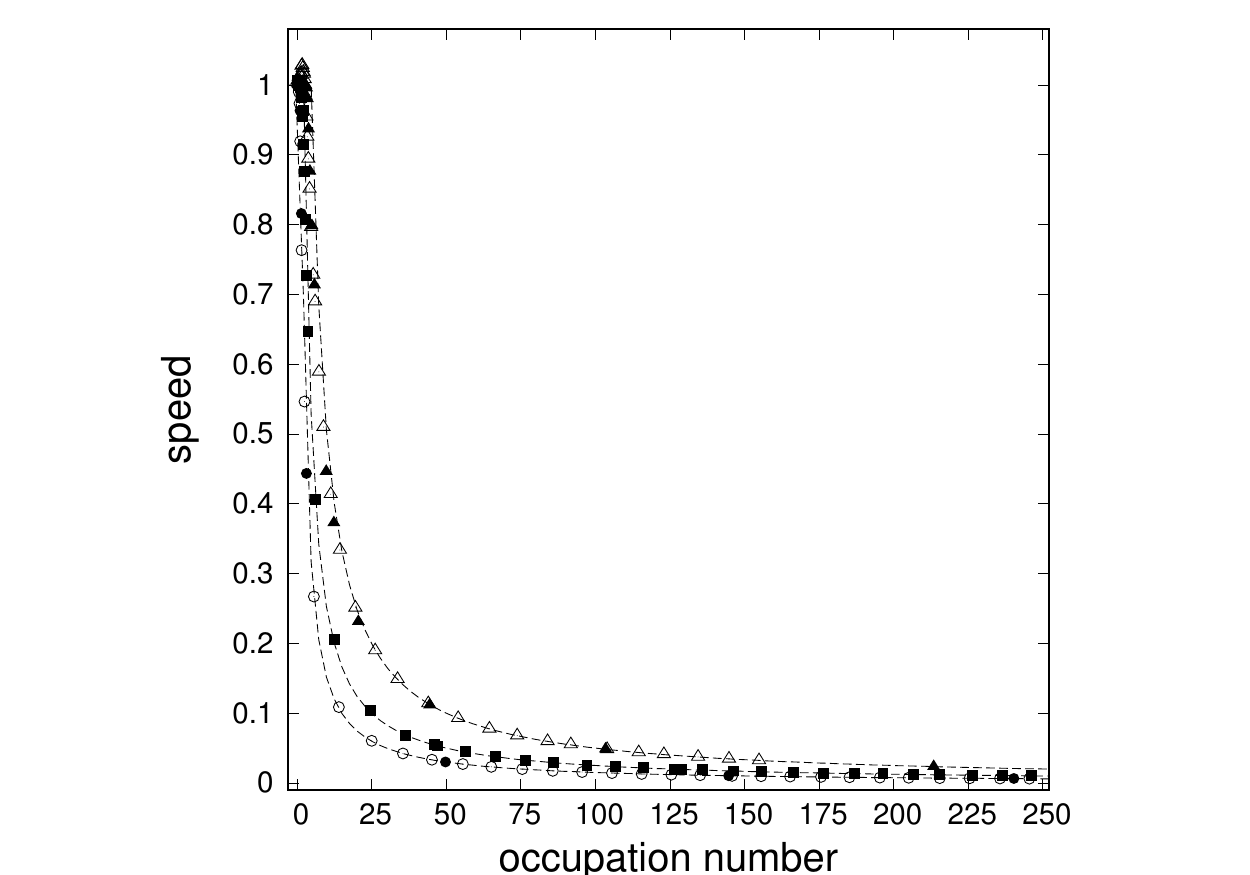}
\end{picture}
\caption{Speed (stationary current divided times the 
stationary occupation number) at the defect site vs.\ the stationary 
occupation number at the same site. 
Open and solid symbols are the Monte Carlo prediction 
for $L=50$ and $L=500$, respectively. 
All simulations refer to the case $p=1$. 
Triangles refer to the case $T=3$ and $c=5$ with 
$\rho$ ranging from $0.2$ to $6.0$.
Squares refer to the case $T=6$ and $c=2.5$ with 
$\rho$ ranging from $0.2$ to $3.5$.
Circles refer to the case $T=4$ and $c=1.5$ with 
$\rho$ ranging from $0.2$ to $2.5$.
The dashed lines are the theoretical prediction: the speed is $1$ in the fluid 
phase and $c/m_1$ in the condensed phase, which correspond in the 
picture to the regions $m_1<c$ and $m_1>c$, respectively. 
The left panel is a magnification of the right panel for low values of the 
stationary occupation number at the defect site. 
}
\label{fig:fd000}
\end{figure}

At the defect site we have to distinguish between the fluid and the 
condensed state. In the former, see Table~\ref{t:mod000} for $p=1$, 
the speed is one, whereas in the latter it is $c/m_1$ (note we 
are considering a large volume situation, but we are not considering 
the thermodynamic limit, indeed, $m_1$ would diverge in the 
condensed state). In other words, 
since in the condensed state the current is constant at the defect site, 
we have that there the local speed decreases as the inverse 
of the mean occupation number.
This behavior is observed 
experimentally in\cite[Figure~4]{DH2005}, where 
the authors perform an experiment with walkers moving 
on a lane and forced to pass through a corridor with reduced 
width with respect to the main lane. The authors measure the 
speed of the pedestrians at the bottleneck as a function of their 
density at the same spot. They find precisely the same behavior we report in Fig.~\ref{fig:fd000} for $c=1.5$.

Figure~\ref{fig:fd000} is obtained by plotting the stationary 
speed at the defect site as a function of the stationary 
occupation number at the same site. Each curve in the plot is 
obtained for a fixed value of the saturated rate $c$. We consider 
the cases $c=1.5,2.5,5$ (see the figure caption for more details). 
For each value of $c$, we obtain the 
different stationary states plotted in the graph (circles, squares, and 
triangles),
by varying the total density $\rho$. According to the results 
in Table~\ref{t:mod000}, in this way we obtain fluid stationary states 
with $m_1=\rho$ for $\rho$ smaller that $c$ and condensed 
states with $m_1=(\rho-c)L+c$ for $\rho$ larger than $c$. 
The dashed lines represent the theoretical prediction: speed 
equal to $1$ in the 
fluid state and to $c/m_1$ in the condensed one. As for the simulations 
discussed in figures~\ref{fig:thr1}--\ref{fig:thr3} the match between the 
theoretical and the Monte Carlo results is strikingly good. 

It is important to remark that our results for $c=1.5$ 
reproduce very well the 
experimental behavior illustrated in\cite[Figure~4 left--top panel]{DH2005} 
for large local densities. 
We use the value $c=1.5$ since in\cite[Figure~4 left--bottom panel]{DH2005} 
it is shown that 
in this density regime the current is approximatively equal 
to $1.5$. This value for $c$ can be also guessed by looking 
at the picture of the experimental setup shown 
in\cite[Figure~2]{DH2005}. Indeed the corridor is so narrow 
that walkers can pass through one or two at a time. 
Since our model predicts very well the experimental behavior, 
we conclude that the structure of the 
experimental fundamental diagram is essentially 
due to the fact that the stationary current in the corridor 
does not depend on the number of pedestrians approaching it.
At low local densities, experimental results depart from the 
inverse proportionality behavior with respect to the local densities, but 
the abrubt onset of the constant behavior that we observe in our model 
is not seen. This is quite natural, since in our model the threshold 
effect is sharp, whereas we expect that in crowd evacuation 
experiments the capacity 
reduction becomes effective in a sort of mild continuous fashion.
 
It is also worth mentioning that in\cite{SSL2006} pedestrian fundamental 
diagrams are studied in the framework of a social force model with an hard core 
parameter reflecting the size of the pedestrians. 
The fundamental diagram in\cite[Figures~1--2]{SSL2006} show some 
similarities with those we found in Fig.~\ref{fig:fd000}, although 
the setup is completely different; in particular no bottleneck 
effect is included in their model. 
The diagrams show the same saturation 
effect at low densities, meaning that in that regime pedestrians
move freely with their own velocity and are not influenced by 
other walkers. 

Results related to the ones we discuss in this paper can be found 
in\cite{TTN2001}, where the authors
study numerically a two--dimensional lattice model, similar 
to the one considered in\cite{CMKS2016}, in which 
particles perform a biased random walk in a square with 
an exclusion rule. Particles preferentially move to the right 
and a sort of bottleneck constraint the flow in the middle 
of the lattice. The current is measured as a funtion of the 
rate at which particles enter the left boundary and profiles 
similar to those that we plotted in figures~\ref{fig:thr2} and \ref{fig:thr3}
(right panel) 
are found. In other words their two--dimensional model shows 
a behavior similar to the one that we proved for our one--dimensional 
case: induced by a large left boudary entry rate a sort of condesed 
stationary state onset is observed numerically and in such a state 
the pedestran current becomes constant. 

By letting the
threshold scale with $L$, in the thermodynamic limit our model only gives rise
to fluid regimes. The trapping state is indeed prevented by the absence of
large enough fluctuations in the number of particles at the bottleneck
(pedestrian density waves\cite{SKKKRS2009}).

%\section{Conclusions}
%\label{s:conclusioni} 
%%\par\noindent

%%\def\acknowledgementname{Funding}
%\textbf{Acknowledgements}
%The authors thank E.\ Presutti, A.\ De Masi, B.\ Scoppola, D.\ Gabrielli and 
%C.\ Landim for many discussions and clarifying remarks. 

%\bibliographystyle{...}
%\bibliography{...}

\end{document}